\documentclass[12pt]{article}
\usepackage{graphics}
\usepackage{epsf}
\usepackage{amsfonts}
\usepackage{amssymb}

\catcode `@=11 \@addtoreset{equation}{section} \catcode `@=12

\def\tsector#1#2{\ {\scriptstyle #1}\hskip 1mm
\mathop{\opensquare}\limits_{\lower
1mm\hbox{$\scriptstyle#2$}}^\sim\hskip 1mm}
\def\appendix{{\newpage\section*{Appendix}}\let\appendix\section

{\setcounter{section}{0}
\gdef\thesection{\Alph{section}}}\section}

\def\){\right)}
\def\({\left( }
\def\]{\right] }
\def\[{\left[ }

\def\half{{1\over 2}}

\newcommand{\be}{\begin{equation}}
\newcommand{\ee}{\end{equation}}
\newcommand{\ba}{\begin{eqnarray}}
\newcommand{\ea}{\end{eqnarray}}
\newcommand{\no}{\nonumber \\}

\newcommand{\p}{\partial}

\def\C{{\mathbb{C}}}
\def\Z{{\mathbb{Z}}}
\begin{document}
\title {{\small \hfill SLAC-PUB-10090  }~~~~~~~\\
Chiral rings and GSO projection in Orbifolds}
\author{ Sunggeun Lee$^1$ and Sang-Jin Sin$^{1,2}$ \\
\small \sl
$^1$Department of Physics, Hanyang University, Seoul, 133-791, Korea,
\\ \small \sl
$^2$Stanford Linear Accelerator Center, Stanford University, Stanford
CA 94305 \footnote{Work supported partially by the department of Energy
under contract number DE-AC03-76SF005515. }
}

\maketitle
\begin{abstract}
The GSO projection in the twisted sector of orbifold background
is sometimes subtle and incompatible descriptions are found in
literatures. Here, from the equivalence of partition functions
in NSR and GS formalisms, we give a simple rule of GSO projection for
the chiral rings of string theory in $\C^r/\Z_n$, $r=1,2,3$. Necessary
constructions of chiral rings are given by explicit mode analysis.
\end{abstract}

\newpage
\section{Introduction}
Recent study of closed string tachyon
analysis\cite{aps,vafa,hkmm,takayanagi,sin2} raised the interests in the
string theory in non-compact orbifold backgrounds.
The essential ingredient in either chiral ring technique\cite{hkmm} or
mirror symmetry approach\cite{vafa,dv,sin3} is the
world sheet ${\cal N}=2$ supersymmetry of NSR
formalism, which in turn requires a precise understanding of the GSO projection. However, the description of this basic ingredient is rather subtle and the subtlety comes from following reason: in untwisted string theory, the
criteria is the spacetime supersymetry which fixes a $\Z_2$ projection unambiguously. However, in orbifold theory, the
spacetime SUSY is broken even after we impose a GSO projection and the imposition of SUSY in untwisted sector does not fix the action of $\Z_2$ projection in the twisted sector uniquely, in the sense that there are many ways of defining $\Z_2$ operartion. Hence the choice of GSO is largely arbitrary and  in fact it is not hard to find that descriptions in literatures are partially incompatible with one another. Furthermore it is not clear whether such GSO projection rules give string theories equivalent to those in Green-Schwarz formalism. 

 In this paper, we explicitly work out the chiral rings and GSO
projection rules on them for non-compact orbifolds $\C^r/\Z_n$, $r=3$ such that it guarantee the equivalence of the NSR and Green-Schwarz formalisms. We find that 
GS-NSR equivalence uniquely fixes a GSO for all twisted sectors and gives us a {simple} rule for the GSO projection. For $r=1,2$, the same GSO rule was  given in the paper \cite{sin2}.

The rest paper is as follows:
After constructing the chiral ring by mode analysis in section 2, we
derive the GSO rule in section 3 by looking at the low temperature limits
of partition function.
We derive partition functions of NSR formalism starting from that of
Green-Schwarz one using the Riemann's theta identities.
The spectrum analysis from this partition function gives the surviving
condition of individual energy values, which we identify as the rule of
the GSO projection.

\section{Construction of Chiral Rings of $\C^r/\Z_n$ }
The purpose of this section is to explicitly construct the chiral rings 
of orbifold theories\cite{dixon}, which is essential both in language and in physical
interpretation of section 3. We use mode analysis and rewrite the result
in terms of monomials of mirror Landau-Ginzburg picture \cite{vafa}, whose review is not included here.   

\subsection{$\C^1/\Z_n$}
The Energy momentum tensor of the NSR string on the cone $\C^1/\Z_n$ is
\be
T=-\p X {\p}X^* + \half \psi^*\p \psi + \half \psi \p \psi^*,\ee
where $X=X^1+iX^2, \; X^*=X^1-iX^2$ and $\psi$ and $\psi^*$ are Weyl
fermions which are
conjugate to each other with respect to the target space complex
structure.
All the fields appearing here describe worldsheet left movers. We
denote the corresponding worldsheet complex conjugate by barred fields:
${\bar X}, {\bar X^*}, {\bar \psi},
{\bar \psi^*}$.
The ${\cal N}=2$ world sheet SCFT algebra is generated by
$T$, $G^+=\psi^*\p X$, $G^-=\psi\p X^*$ and $J=\psi^*\psi$.
The orbifold symmetry group
\be \Z_n=\{g^l| l=0,1,2,\cdots,n-1, {\rm with} \;\; g^n=1\}
\ee act on the fields in NS sector by
\ba
g\cdot X(\sigma+2\pi,\tau) &=& e^{2\pi i k/n}g\cdot X(\sigma,\tau),\;\;
\no
g\cdot X^*(\sigma+2\pi,\tau) &=&   e^{-2\pi i k/n}g\cdot
X^*(\sigma,\tau),\no
g\cdot \psi(\sigma+2\pi,\tau) &=& - e^{2\pi i k/n}g\cdot \psi
(\sigma,\tau),\;\; \no
g\cdot \psi^*(\sigma+2\pi,\tau) &=&  - e^{-2\pi i k/n}g\cdot
\psi^*(\sigma,\tau).
\ea
The mode expansions of the the fields in the conformal plane are given by
\ba
\p X(z)=& \sum_{n\in\Z} \alpha_{n+a}/z^{n+1+a}, \;\no
\p X^*(z)=& \sum_{n\in\Z} \alpha^*_{n-a}/z^{n+1-a},\; \no
\psi(z)=& \sum_{r\in\Z+\half} \psi_{r+a}/z^{r+\half+a},\;\no
\psi^*(z)=& \sum_{r\in\Z+\half} \psi^*_{r-a}/z^{r+\half-a} ,
\label{modes}
\ea
where $a=k/n$.
The quantization conditions are:
\be
[\alpha_{n+a},\alpha^*_{-m-a}]=(n+a)\delta_{m,n}, \quad
\{\psi_{r+a},\psi^*_{-s-a}\}=\delta_{r,s}.\ee
Hence the conjugate variables are given by
\ba
\alpha^\dagger_{n+a}= \alpha^*_{-n-a}, &\; (\alpha^*_{n-a})^\dagger=
\alpha_{-n+a},\;
\no
\psi^\dagger_{r+a}=\psi^*_{-r-a},&
(\psi^*_{r-a})^\dagger=\psi_{-r+a}.
\ea

The vacuum is defined as a state that is annihilated by all positive
modes.
Notice that as $a$ grows greater than $\half$, $\psi_{-\half+a}$
($\psi^*_{\half-a}$)changes
from a creation(annihilation) to an annihilation (creation) operator.
The (left mode) hamiltonian of the orbifolded complex plane is
\be
H_L=\half\sum \[\alpha^*_{-n-a}\alpha_{n+a}+
\alpha_{-n+a}\alpha^*_{n-a}\] +
\half\sum_{r\in\Z+\half}\[(r+a)\psi^*_{-r-a}\psi_{r+a}+
(r-a)\psi_{-r+a}\psi^*_{n-a}\] . \ee
The contribution of the left modes to the zero energy is
\be E_0^L = \half\sum_{n=0}^{\infty}(n+a)+\half\sum_{n=1}^{\infty}(n-a)
-\half\sum_{n=0}^\infty(n+\half+a)-\half\sum_{n=0}^{\infty}(\half+n-a).\ee
If we define
\be
f(a)=\sum_{n=0}^{\infty}(n+a)=1/24-(a-1/2)^2/2,\ee
then $f(a)=f(1-a)$ and
$f(a+1/2)=f(-a+1/2)$ so that the above sum gives $E_0^L=a/2-1/8$.
Embedding
the cone to the string theory to make the target space $\C/\Z_n\times
R^{7,1}$,
we need to add the zero energy fluctuation of the 6 transverse flat
space,
$ 6\times (-1/24)(1+1/2)=-3/8$ to the zero point energy, which finally
become
\be
E_0^L=\half(a-1), \quad for\quad 0<a<\half.
\ee
If $1/2<a<1$, then $(a-\half)\psi^*_{\half-a}\psi_{-\half+a}$ should be
added to the
normal ordered Hamiltonian while
$(\half-a)\psi_{-\half+a}\psi^*_{\half-a}$ should be removed from it.
Therefore the zero point energy should be modified to be
\be
E_0^L=\half\(a-1\)- \half\[-\(\half-a\)+\(a-\half\)\]=-\half a, \quad for
\quad \half<a<1.
\ee
From this we can identify the weight and charge of twisted ground states
using
\be
E_0^L=\Delta-1/2, \quad {\rm and}\quad q=\pm 2\Delta,
\ee
where we take + if the the ground state is a chiral state and $-$ if
it is anti-chiral
state.

We now construct next level chiral and anti-chiral primary states
by applying the creation operator.
\ba
G^+_{-\half}&=&\sum \psi^*_{-n-\half-a} \alpha_{n+a}=
\psi^*_{\half-a}\alpha_{-1+a} +\cdots
,\;\no
G^-_{-\half}&=&\sum \psi_{-n-\half+a} \alpha^*_{n-a} =
\psi_{-\half+a}\alpha^*_{-a} +\cdots.
\ea
Notice that for $0<a<\half$, $\psi^*_{\half-a}|0> = 0$, hence
\be G^+_{-\half}|0>=0\ee
so that $|0>$ is a chiral state. It has a weight $a/2$ and a R-charge
$a$, so
that the local ring element of LG theory corresponding to $|0>_a$ can be
identified as $u^k$:
\be
|0>_a \sim u^k.\ee
The first excited state is $\psi_{a-\half}|0>$ which is
annihilated by $G^-_{-\half}$:
\be
G^-_{-\half} \psi_{a-\half}|0> =0 .\ee
Therefore it is an anti-chiral state.
Its weight is $\half(1-a)$ and charge is $a-1$, hence it
corresponds to ${\bar u}^{n-k}$:
\be
\psi_{a-\half}|0> \sim {\bar u}^{n-k}.\ee

For $\half<a<1$, $\psi_{a-\half}|0> = 0$, hence
$G^-_{-\half}|0>=0$ so that $|0>$ is an anti-chiral state. It has
weight $\half(1-a)$ and charge $a-1$, hence the corresponding
local ring element is ${\bar u}^{n-k}$. The first excited state is
$\psi^*_{\half-a}|0>$ which is annihilated by $G^+_{-\half}$
therefore it is a chiral state. Its weight is $a/2$ and the charge
is $a$ hence the corresponding local ring element is again $u^k$.
Using the weight and charge relation for chiral and anti-chiral
states, we see that $\psi^*$ has $+1$ charge and $\psi$ has $-1$
charge.

So far we have worked out the first twisted sector for arbitrary
generator $k$.
For the $j$-th twisted sector, we can easily extend the above
identifications
by observing that $a$ is the fractional part of $jk/n$;
\be a=\{jk/n\} .\ee
The result is that for all chiral operators, the local ring
elements are given by $u^{ n\{jk/n\} }$ and for the anti-chiral
operators they are given by ${\bar u}^{ n-n\{jk/n\} }$. In both
cases $j$ runs from 1 to $n-1$ for twisted sectors.
It is worthwhile to observe that
\be
n(1-\{jk/n\})=n\{j(n-k)/n\},\ee
so that the generator of the anti-chiral ring is ${\bar u}^{n-k}$, while
that of chiral ring is ${u}^{k}$.
Since it is the building block for
the results in higher dimensional theories, we tabulate the above results
in Table \ref{ring}.
\begin{table}
 \centering
\begin{tabular}{|c|c|c|c|} \hline
Region & vacuum & annihilator & creators \\ \hline
$0<a<\half$ & $|0>\sim u^{na}$ & $\psi^*_{\half-a}$ &
$\psi_{a-\half}|0>\sim {\bar u}^{n-na}$ \\ \hline
$\half<a<1$ & $|0>\sim {\bar u}^{n-na}$ & $\psi_{a-\half}$ &
$\psi^*_{\half-a} |0>\sim { u}^{na}$\\ \hline
\end{tabular}
 \caption{\scriptsize
 Twisted vacuum and first excited states. The chirality is equal to the
holomorphic
 structure of the target space, i.e, chiral(anti-chiral) states
correspond to monomial of $u$ ($\bar u$). }
 \label{ring}
\end{table}

What about the case $a=-k/n<0$ The answer can be read
off from what we already have by noticing that above structure is
periodic
in $a$ with period 1, because we should shift the mode if $a$ is bigger
than 1.
The effect of $a\to -a$ amounts to exchanging the role of $\psi$ and
$\psi^*$.
Therefore in this case, the local ring
elements of LG dual are given by $u^{n-n\{jk/n\} }$ and for the
anti-chiral
operators they are given by ${\bar u}^{n\{jk/n\} }$.

\begin{figure}[htbp2]
\epsfysize=6cm
\centerline{\epsfbox{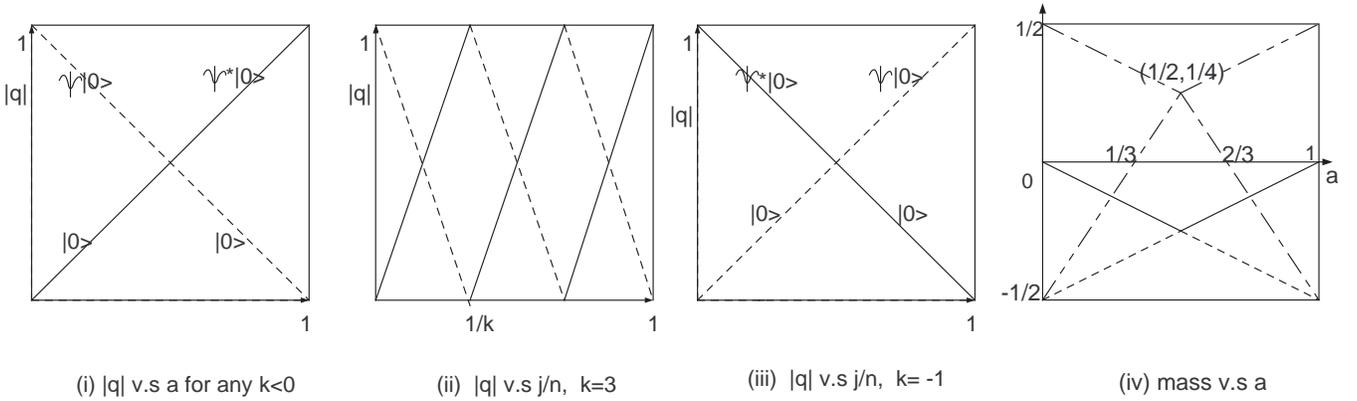}}
\caption{\scriptsize Spectrum versus twists in $\C^1/\Z_n$:
(i) $2\Delta=|q|$ v.s $a=\{jk/n\}$ for any $k>0$. The states on solid
lines are chiral while those on the dotted lines are anti-chiral. (ii)
$|q|$ v.s $j/n$ for k=3,
(iii) $|q|$ v.s $j/n$ for $k=-1$. For $k<0$, the role of chiral and
anti-chiral states are interchanged.
(iv) All possible Tachyom mass v.s $a$. Dotted lines are for the
(twisted) vacuum,
solid lines are for worldsheet fermion excitations, the rests are for
scalar excitations.
 Notice that the lowest tachyon mass is always generated by worldsheet
fermion.}
 \label{Fig0}
\end{figure}

The first three graphs in Fig.\ref{Fig0} show the weight versus twist $a$
for the various cases.
The charge can be read off by the $q=\pm 2\Delta$ rule. We are interested
in $2\Delta$ since left and
right moving parts contribute the same to the masses of the states
represented by these polynomials.
The last figure in Fig.\ref{Fig0} is mass spectrum $\frac{1}{4}\alpha'
M^2=E_0^L$
as a function of the twist $a$ for {\it all} possible tachyons
including the scalar excitations:
\ba
 \alpha^*_{-a} |0 \rangle :\; E_0^L =& (3a-1)/2, \; {\rm for} \;
&0<a<\half \no
=& {a}/{2}, \;\;\; {\rm for} \; &
\half<a<1 \no
 \alpha_{-(1-a)}|0 \rangle : \; E_0^L =& (1-a)/, \; {\rm for} \; &
0<a<\half \no
=& {(2-3a)}/{2}, \; {\rm for} \; &
\half<a<1 .
\ea
These scalar excitations $\alpha^*_{-a} |0 \rangle, \alpha_{-(1-a)}|0
\rangle $
are tachyons if $0<a<1/3$, $2/3<a<1$ respectively. They
can not be characterized as chiral or anti-chiral states.
Furthermore it never gives the lowest tachyon mass, hence we will not
pay
much attention afterward.

\subsection{$\C^2/\Z_n$}
Now we extend the result of previous section to $\C^2/\Z_n$ case,
which is our main interests. We introduce two
sets of (bosonic and fermionic) complex fields
$X^{(1)},\psi^{(1)},\psi^{*(1)};$
$X^{(2)},X^{*(2)},\psi^{(2)},\psi^{*(2)}$
and specify how the orbifold group $\Z_n$ is acting on each set
of fields. The group action is the same as before except that
$\Z_n$ can act on first and second set of fields with different
generators $k_1$ and $k_2$. For example, in the first twisted sector,
\be
g\cdot X^{(j)}(\sigma+2\pi,\tau)=e^{2\pi i k_j/n}g\cdot
X^{(j)}(\sigma,\tau),
\quad for \quad j=1,2 . \ee
Since three parameter $n,k_1,k_2$ fix an $\C^2/\Z_n$ orbifold theory
completely,
we use notation $n(k_1,k_2)$ to denote it.

Let $a_i=k_i/n$ as before. For $0<a_i<\half$, the zero energy
fluctuation can be calculated as
\be
E_0^L=\(\half a_1-\frac{1}{8} \)
+\(\half a_2-\frac{1}{8} \) -\frac{1}{24}
\(1+\half\)\times 4= \half\(a_1+a_2-1\).
\ee
Therefore the weight of twisted vacuum is given by \be
\Delta_0=\half(a_1+a_2). \ee
We define
\be G^+=G_1^++G_2^+,\ee
where
$G_i^+=\psi^{*(i)}\p X^{(i)}$.
For abbreviation, we use following notations;
\be
\psi_i:=\psi^{(i)}_{a_i-\half}\quad {\rm and}\;\;
\psi_i^*:=\psi^{*(i)}_{\half-a_i}.\ee
 Then
for $a_1<\half,a_2<\half$, we have $\psi^*_1|0>=0$ and $\psi_2^*|0>=0$,
which gives
 $G^+_{-\half}|0>=0$ so that the twisted vacuum is a chiral state,
whose associated local ring element is identified:
\be
|0>\sim u_1^{n\{jk_1/n\}}u_2^{n\{jk_2/n\}}.\ee
Both $\psi_1,\psi_2$ are creation operators and
$G^-_{-\half}\psi_1\psi_2|0>=0$, so that $\psi_1\psi_2|0>$ is an
anti-chiral
state. By considering weight and charge,  corresponding monomial
is found to be
\be \psi_1\psi_2|0> \sim
{\bar u_1}^{n-n\{jk_1/n\}}{\bar u_2}^{n-n\{jk_2/n\}}.\ee
So far, $ \psi^*_i|0>$'s are neither chiral($c$) nor anti-chiral($a$).
One can work out other three cases in similar fashion. We
summarize the result in the Table \ref{modeNmonomial}.
\begin{table}
 \centering
\begin{tabular}{|c|c|c|c|c|c|} \hline
($a_1-\half$,$a_2-\half$) & $b$ & $b^\dagger$ & chiral state &
anti-chiral state & neither \\ \hline
$--$ & $\psi^*_1,\psi_2^*$ & $\psi_1,\psi_2$ & $|0>$ & $\psi_1\psi_2|0>$
& $\psi_1|0>\sim {\bar u_1}^{n-na_1}u_2^{na_2}$ \\
~&~&~&$\sim u_1^{na_1}u_2^{na_2}$ & $\sim {\bar u_1}^{n-na_1}{\bar
u_2}^{n-na_2}$ & $\psi_2|0>\sim { u_1}^{na_1}{\bar u}_2^{n-na_2}$ \\
\hline
$-+$ & $\psi^*_1,\psi_2$ & $\psi_1,\psi^*_2$ & $\psi_2^*|0> $ &
$\psi_1|0> $ & $|0>\sim { u_1}^{na_1}{\bar u}_2^{n-na_2}$\\
~&~&~&$\sim u_1^{na_1}u_2^{na_2}$ & $\sim {\bar u_1}^{n-na_1}{\bar
u_2}^{n-na_2}$ & $\psi_1\psi_2^*|0>\sim {\bar u_1}^{n-na_1}u_2^{na_2}$
\\ \hline
$+-$ & $\psi_1,\psi_2^*$ &$\psi^*_1,\psi_2$ & $\psi_1^*|0> $ &
$\psi_2|0> $ & $|0>\sim {\bar u_1}^{n-na_1}u_2^{na_2} $ \\
~&~&~&$\sim u_1^{na_1}u_2^{na_2}$ & $\sim {\bar u_1}^{n-na_1}{\bar
u_2}^{n-na_2}$ & $\psi^*_1\psi_2|0> \sim { u_1}^{na_1}{\bar
u}_2^{n-na_2}$ \\ \hline
$++$ & $\psi_1,\psi_2$ & $\psi^*_1,\psi^*_2$ & $\psi_1^*\psi_2^*|0> $ &
$|0> $ & $\psi^*_1|0>\sim { u_1}^{na_1}{\bar u}_2^{n-na_2}$ \\
~&~&~&$\sim u_1^{na_1}u_2^{na_2}$ & $\sim {\bar u_1}^{n-na_1}{\bar
u_2}^{n-na_2}$ & $ \psi^*_2|0> \sim {\bar u_1}^{n-na_1}u_2^{na_2} $ \\ 
\hline
\end{tabular}
\caption{\scriptsize Oscillator and monomial representations of chiral
and anti-chiral
rings. $+-$ means ($a_1-\half)>0,(a_2-\half)<0$. }\label{modeNmonomial}
\end{table}

Notice that (anti-)chiral states in
different parameter ranges have different oscillator
representations but have the same polynomial expressions as local
ring elements.

When some of $a_i<0$, one can get the similar result by exchanging
the role of $\psi$ and $\psi^*$, and $u_i$ and ${\bar u}_i$. As a
result, for the factor with the negative twist $a_i=-\{jk_i/n\}$,
we need to use $u_i^{n-n\{jk_i/n\}}$ for the chiral states and
${\bar u_i}^{n\{jk_i/n\}}$ for the anti-chiral states, while for
the factor with the positive twist $\{jk_i/n\}$, we need to use
$u_i^{n\{jk_i/n\}}$ for the chiral states and ${\bar
u_i}^{n\{jk_i/n\}}$ for the anti-chiral states. For example: if
only $a_2$ is negative, the chiral states are associated with
$u_1^{n\{jk_1/n\}}u_2^{n-n\{jk_2/n\}}$, while the anti-chiral
states are associated with $ {\bar u_1}^{n-n\{jk_1/n\}}{\bar
u_2}^{n\{jk_2/n\}}$. We summarize the result in the Table
\ref{sign}.

\begin{table}
 \centering
\begin{tabular}{|c||c|c||c|c|} \hline
 $(a_1,a_2)$ & $c$-ring & $2\Delta$ & $a$-ring & $2\Delta$
 \\ \hline \hline
$ ++$ & $u_1^{na_1}u_2^{na_2}$ & $a_1+a_2$ & ${\bar u_1}^{n(1-a_1)}{\bar
u_2}^{n(1-a_2)}$ & $2-a_1-a_2$
 \\ \hline
$ +-$ & $ u_1^{na_1}u_2^{n(1-|a_2|)}$ & $ a_1-|a_2|+1$ & $ {\bar
u_1}^{n(1-a_1)}{\bar u_2}^{n|a_2|}$ & $ 1-a_1+|a_2|$
 \\ \hline
$ -+$ & $ u_1^{n(1-|a_1|)}u_2^{na_2}$ & $ 1-|a_1|+a_2$ & ${\bar
u_1}^{n|a_1|}{\bar u_2}^{n(1-a_2)}$ & $ |a_1|-a_2+1$
 \\ \hline
$--$ & $ u_1^{n(1-|a_1|)}u_2^{n(1-|a_2|)}$ & $ 2-|a_1|-|a_2|$ & ${\bar
u_1}^{n|a_1|}{\bar u_2}^{n|a_2|}$ & $ |a_1|+|a_2|$
 \\ \hline
\end{tabular}
 \caption{\scriptsize Monomial basis of chiral and anti-chiral rings and
their weights when some of $a_i$ is negative.
 $+-$ means $a_1>0,a_2<0$. The R-charges can be read off
 by the rule $q=\pm 2\Delta$. }\label{sign}
\end{table}

\subsection{$\C^3/\Z_n$}

Now let $a_i =k_i /n$ as before with $i=1,2,3$ and consider first $0<
a_i <{1\over 2}$.
The zero energy fluctuation can be calculated as
\be
E_0^L=\sum_i ({1\over 2}a_i -{1\over 8})-{1\over 24}(1+{1\over 2})\times
2
={1\over 2}(a_1 + a_2 + a_3 -1).
\ee
Hence the weight of the twisted vacuum can be read off as before
\be
\Delta_0 = {1\over 2}(a_1 +a_2 +a_3).
\ee
We define
\be
G^+ =G^+_1 +G^+_2 +G^+_3,
\ee
where $G^+_i =\psi^{*(i)}\partial X^{(i)}$. For abbreviation we
use the following notations
\be
\psi_i:=\psi^{(i)}_{a_i -{1\over 2}}\;\;\;\;\;
\psi^*_i:=\psi^{*(i)}_{{1\over 2}-a_i}.
\ee
Suppose $a_i <{1\over 2}$ then
\be
\psi^*_i |0> =0 \quad i.e.\quad G^+_{-{1\over 2}}|0>=0.
\ee
Thus the twisted vacuum is a chiral state, in other words
\be
|0> \sim u_1^{n\{ {jk_1 /n} \} }u_2^{n\{jk_2/n\} }u_3^{n\{jk_3/n\}},
\ee
while
\be
G^-_{-{1\over 2}} \psi_1 \psi_2 \psi_3 |0>=0,
\ee
meaning
\be
\psi_1\psi_2 \psi_3 |0> \sim \bar{u}_1^{n-n\{ jk_1 / n \} }
\bar{u}_2^{n-n\{jk_2/n\}} \bar{u}_3^{ n-n\{jk_3/n\}}.
\ee
is an anti-chiral state. We summarize this in Table \ref{t1} and for
$a_i <0$
in Table \ref{t2}.

\begin{table}
\centering
 \begin{tabular}{|c|c|c|c|c|} \hline
$(a_1-{1\over 2},a_2-{1\over 2},a_3-{1\over 2})$ & $b$ & $b^\dagger$
& chiral & anti-chiral \\ \hline
$(-,-,-)$ & $\psi^*_1,\psi^*_2,\psi^*_3$ &$ \psi_1,\psi_2,\psi_3$ & $|0>
$ &$ \psi_1 \psi_2 \psi_3|0> $
\\
~&~&~&$\sim u_1^{na_1}u_2^{na_2}u_3^{na_3}$&$\sim
\bar{u}_1^{n-na_1}\bar{u}_2^{n-na_2} \bar{u}_3^{n-na_3}$ \\ \hline
$(-,+,-)$ &$ \psi^*_1,\psi_2,\psi^*_3$ &
$ \psi_1,\psi^*_2, \psi_3 $& $\psi^*_2|0> $ & $\psi_1|0> $ \\
~&~&~& $ \sim u_1^{na_1}u_2^{na_3} u_3^{na_3}  $& $ \sim
\bar{u}_1^{n-na_1}
\bar{u}_2^{n-na_2}\bar{u}_3^{n-na_3}  $ \\ \hline
$(-,-,+)$ & $\psi^*_1, \psi^*_2, \psi_3$ &$ \psi_1, \psi_2, \psi^*_3$ &
$\psi^*_3|0>$ &$ \psi_1 \psi_2|0> $ \\
~&~&~& $ \sim u_1^{na_1}u_2^{na_2}u_3^{na_3} $ &$  \sim
\bar{u}_1^{n-na_1} \bar{u}_2^{n-nq_2} u_3^{n-na_3}    $ \\ \hline
$(+,-,-)$ &$ \psi_1, \psi^*_2, \psi^*_3$ & $\psi^*_1,\psi_2, \psi_3$ &
$\psi^*_1|0>$ & $\psi_2 \psi_3 |0> $ \\
~&~&~& $ \sim u_1^{na_1}u_2^{na_2}u_3^{na_3}   $ & $  \sim
\bar{u}_1^{n-na_1}\bar{u}_2^{n-na_2}\bar{u}_3^{n-na_3}    $ \\ \hline
$(+,+,-)$ &$ \psi_1, \psi_2, \psi^*_3$ &$ \psi^*_1, \psi^*_2, \psi_3$ &
$\psi^*_1 \psi^*_2|0> $ &
$\psi_3|0> $ \\
~&~&~& $ \sim u_1^{na_1}u_2^{na_2}u_3^{na_3} $&
$ \sim \bar{u}_1^{n-na_1} \bar{u}_2^{n-na_2}\bar{u}_3^{n-na_3}  $  \\
\hline
$(+,-,+)$ & $\psi_1,\psi^*_2,\psi_3$ & $\psi^*_1,\psi_2,\psi^*_3$ &
$\psi^*_1, \psi^*_3 |0> $ &
$\psi_2 |0> $
\\
~&~&~&$\sim u_1^{na_1}u_2^{na_2}u_3^{na_3} $ &
$ \sim \bar{u}_1^{n-na_1} \bar{u}_2^{n-na_2} \bar{u}_3^{n-na_3} $ \\
\hline
$(-,+,+)$ & $\psi^*_1, \psi_2, \psi_3$ & $\psi_1, \psi^*_2, \psi^*_3$ &
$\psi^*_2 \psi^*_3|0> $ &
$\psi_1 |0> $ \\
~&~&~& $ \sim u_1^{na_1}u_2^{na_2}u_3^{na_3} $ &
$ \sim \bar{u}_1^{n-na_1}\bar{u}_2^{n-na_2} \bar{u}_3^{n-na_3} $ \\
\hline
$(+,+,+)$ & $\psi_1, \psi_2, \psi_3 $& $\psi^*_1, \psi^*_2, \psi^*_3$ &
$\psi^*_1 \psi^*_2 \psi^*_3 |0> $ &
$|0> $ \\
~&~&~& $ \sim u_1^{na_1}u_2^{na_2}u_3^{na_3} $ &
$ \sim \bar{u}_1^{n-na_1}\bar{u}_2^{n-na_2}\bar{u}_3^{n-na_3} $ \\
\hline
\end{tabular}
\caption{Oscillator and monomial representations of chita and anti-chiral
rings:
$-(+)$ means $a_i < {1\over 2}(a_i > {1\over 2})$.}\label{t1}
\end{table}

\begin{table}
\centering
\begin{tabular}{|c|c|c|c|c|}\hline
$a_1$, $a_2$, $a_3$ & $c$-ring & 2 $\Delta$ &$a$-ring & 2$\Delta$ \\
\hline
$+,+,+$ & $ u_1^{na_1}u_2^{na_2} u_3^{na_3}$ & $a_1 +a_2+a_3$ &
$\bar{u}_1^{n(1-a_1)}
\bar{u}_2^{n(1-a_2)}\bar{u}_3^{n(1-a_3)}$ & $3-a_1-a_2-a_3$ \\ \hline
$+,-,+$ &$ u_1^{na_1} u_2^{n-n|a_2|} u_3^{na_3} $&$ a_1 +1-|a_2|+a_3$
& $\bar{u}_1^{n(1-a_1)}\bar{u}_2^{n|a_2|}\bar{u}_3^{n(1-a_3)}$ &
$2-a_1+|a_2|-a_3$
\\ \hline
$+,+,-$ & $u_1^{na_1}u_2^{na_2} u_3^{n-n|a_3|} $& $a_1 +a_2 +1-|a_3| $
 &$\bar{u}_1^{n(1-a_1)}\bar{u}_2^{n(1-a_2)}\bar{u}_3^{n|a_3|}$  &
$2-a_1-a_2 +|a_3|$   \\ \hline
$-,+,+$ & $u_1^{n-n|a_1|} u_2^{na_2}u_3^{na_3}$ & $1-|a_1| +a_2 +a_3 $
 & $\bar{u}_1^{n|a_1|}\bar{u}_2^{n(1-a_2)}\bar{u}_3^{n(1-a_3)}$ &
$2+|a_1|-a_2 -a_3$   \\ \hline
$-,-,+$ &$ u_1^{n-n|a_1|}u_2^{n-n|a_2|}u_3^{na_3}$&$2-|a_1|-|a_2|+a_3 $
 &$ \bar{u}_1^{n|a_1|}\bar{u}_2^{n|a_2|}\bar{u}_3^{n(1-a_3)}$   &
$1+|a_1|+|a_2|-a_3$    \\ \hline
$-,+,-$& $ u_1^{n-n|a_1|} u_2^{na_2} u_3^{n-n|a_3|} $&
$2-|a_1|+a_2 -|a_3| $ & $
\bar{u}_1^{n|a_1|}\bar{u}_2^{n(1-a_2)}\bar{u}_3^{n|a_3|}$
 & $ 1+|a_1|-a_2+|a_3|$ \\ \hline
$+,-,-$&$u_1^{na_1}u_2^{n-n|a_2|} u_3^{n-n|a_3|}$&$2+a_1 -|a_2|-|a_3| $
 & $\bar{u}_1^{n(1-a_1)}\bar{u}_2^{n|a_2|}\bar{u}_3^{n|a_3|}$ &
$1-a_1+|a_2+|a_3|$ \\ \hline
$-,-,-$ & $ u_1^{n-n|a_1|}u_2^{n-n|a_2|}u_3^{n-n|a_3|}$ &
$3-|a_1|-|a_2|-|a_3|$ &$
\bar{u}_1^{n|a_1|}\bar{u}_2^{n|a_2|}\bar{u}_3^{n|a_3|}$ &
$|a_1|+|a_2|+|a_3|$ \\ \hline
\end{tabular}
\caption{Monomial basis of chiral and anti-chiral rings and their weights
when some
of $a_i$ are negative.}\label{t2}
\end{table}

\begin{table}
\centering
\begin{tabular}{|c|c|c|c|c|} \hline
$G$ & c-ring & 2 $\Delta$ & $a$-ring & 2$\Delta$ \\ \hline
$G_{ccc}$ & $ u_1^{na_1}u_2^{na_2}u_3^{na_3} $&$ a_1 +a_2 +a_3 $
& $\bar{u}_1^{n(1-a_1)}{\bar u}_2^{n(1-q_2)}\bar{u}_3^{n(1-a_3)}$ &
$3-a_1-a_2-a_3$  
\\ \hline
$G_{cac}$ & $u_1^{na_1}\bar{u}_2^{n-na_2}u_3^{na_3} $ & $ a_1 + 1-a_2 +a
_3 $ &
$\bar{u}_1^{n(1-a_1)}u_2^{na_2}\bar{u}_3^{n(1-a_3)}$ & $2-a_1+a_2-a_3
$
\\ \hline
$G_{cca}$ & $ u_1^{na_1} u_2^{na_2}\bar{u}_3^{n-na_3}$ & $ a_1 + a_2 +
1-a_3 $
& $\bar{u}_1^{n(1-a_1)}\bar{u}_2^{n(1-a_2)}u_3^{na_3}$ & $2-a_1-a_2+a_3$
\\ \hline
$G_{acc} $ & $ \bar{u}_1^{n-na_1} u_2^{na_2}u_3^{na_3}$ & $ 1-a_1 +a_2 +
a_3 $
& $u_1^{na_1}\bar{u}_1^{n(1-a_2)}\bar{u}_3^{n(1-a_3)}$ &$2+a_1-a_2-a_3
$
\\ \hline
$G_{aac}$ & $ \bar{u}_1^{n-na_1}\bar{u}_2^{n-na_2}u_3^{na_3} $&
$ 2-a_1 -a_2 + a_3 $ &$ u_1^{na_1}u_2^{na_2}\bar{u}_3^{n(1-a_3)}$
&$1+a_1+a_2-a_3 $ 
\\ \hline
$G_{aca}$ & $ \bar{u}_1^{n-na_1}u_2^{na_2}\bar{u}_3^{n-na_3}$ &
$2-a_1 +a_2 -a_3$ & $u_1^{na_1}\bar{u}_2^{n(1-a_2)}u_3^{na_3}$ &
$1+a_1-a_2+a_3 $ \\ \hline
$G_{caa}$ & $ u_1^{na_1}\bar{u}_2^{n-na_2}\bar{u}_3^{n-na_3} $ &
$2+a_1 -a_2 -a_3 $ & $\bar{u}_1^{n(1-a_1)}u_2^{na_2}u_3^{na_3} $ &
$1-a_1+a_2+a_3$ \\ \hline
$G_{aaa}$ & $ \bar{u}_1^{n-na_1}\bar{u}_2^{n-na_2}\bar{u}_3^{n-na_3}$ &
$3-a_1 -a_2 -a_3 $ & $u_1^{na_1}u_2^{na_2}u_3^{na_3} $ &
$a_1+a_2+a_3$
\\ \hline

\end{tabular}
\caption{Monomial basis of chiral and anti-chiral rings and their weights
for
varius choices of target space complex structures.} \label{t3}
\end{table}

The discussion on the enhanced (2,2) SUSY can be described completely
parallel way with $\C^2/\Z_n$ case.

\newpage
\section{GSO projection}

\subsection{ Type 0 and type II Orbifold}

Here we discuss when there is bulk tachyons whose presence/absence defines type 0/type II.
 Considering twist operation in Green-Schwarz formalism. We follow the argument in \cite{aps} and generalize it.

First we consider $\C^1/\Z_n$.
Let $g$ be the orbifold action acting on complex plane;
\be
g=e^{2\pi ikJ/n}, \quad k=-n+1, \cdots, n-2, n-1, \label{generator}
\ee
where $J$ is the rotation generator in the complex plane that is
orbifolded.
\be
g^n=(-1)^{kF_s},
\ee
where $F_s$ means spacetime fermion number and we used $J=1/2$ for the
spacetime fermion.
Hence if $k$ is even, then $g^n=1$ and $g$ is a good generator of $\Z_n$
action.
On the other hand, if $k$ is odd, $g^n=(-1)^{F_s}\neq 1$, and $g$ is not
a generator of  $\Z_n$ action.
In fact $g$ is the generator of $\Z_{2n}$ action. The $\Z_{2n}$
projection operator
$P$ projects out all spacetime fermion, since
 \be
 P=\frac{1}{2n}\sum_{l=0}^{2n-1} g^l=\half (1+(-1)^{F_s})\cdot
\sum_{l=0}^{n-1}g^l/n.
 \ee
The consequence is type 0 string where there is no spacetime fermion.
More precisely,
the bulk fermion in untwisted sector is cancelled by those of $n$-th
twisted sector.

In order to get type II string for $k$ odd case, one try to change the
projection operator by
\be
g'=e^{2\pi ikJ/n} (-1)^{-2\pi i J},
\ee
so that $g'^n=(-1)^{(k-n)F_s}$. Notice that under the change $g\to
g'$, it follows that $k\to k-n$
in eq.(\ref{generator}) and $k'=k-n$ is even only if $n$ is odd and the
theory can be a type II \cite{aps}. Notice that after we change $g\to g'$
the theory is changed from $n(k) \to n(k-n)$ and when $n$ is odd, two
are different after GSO projection, though they were the same as a
conformal field theories in NSR formalism. We emphasize that it is not
necessary for $n$ odd for type II if $k$ is even. This is consistent with
\cite{lowe}.

Now we consider $\C^2/\Z_n$.
The twist operator is
\be
g=\exp(2\pi i(k_1J_1+k_2J_2)/n),
\ee
$g^n=(-1)^{F_s(k_1+k_2)}=(-1)^{F_s}$.
Therefore $g$ define a type II theory for $k_1+k_2$ even, and a type 0
theory for $k_1+k_2$ odd.
In order to get a type II theory for $k_1+k_2=odd$, the twist operator
should be modified
to \be
g'=\exp(2\pi i(k_1J_1+k_2J_2)/n) (-1)^{F_s}.
\ee
Since $g'^n=(-1)^{(k_1+k_2-n)F_s}$, we need odd $n$ to get type II
theory in the case of $k_1+k_2$ is odd. 
Again, important notice is that when we twist by $(-1)^{F_s}$, we need
to shift one of $k_i$ to $k_i-n$.

So we get the following {\it lemma: \\ If $k_1+k_2$ is even, the theory
is type II, otherwise it is type 0. }

Since we can choose $k_1=1$ without loss of generality, we need to
consider only $n(1,k)$ and
in this convention, we have type II theory if $k$ is odd and type 0 if
$k$ is even.

$\C^3/\Z_n$ is similar: we have $
g=e^{2\pi i (k_1J_1 +k_2J_2+k_3J_3)/ n}$
which gives
$g^n = (-1)^{(k_1+k_2+k_3)F_s}$.
When $k_1+k_2+k_3$ is $even$ the result is type II while
$odd$ type 0. To be a type II for $odd$ case modify
$g' = e^{2\pi i (k_1 J_1+k_2J_2 +k_3J_3)/ n}(-1)^{F_s}$
which is $g'^n =(-1)^{(k_1+k_2+k_3-n)F_s}$.
Hence we can get type II when $n$ is $odd$ for odd $k_1+k_2+k_3$.

The key notice in all cases is that when we twist by extra
$(-1)^{F_s}$  by $g\to g'$, we are changing $\sum_ik_i \to \sum_ik_i-n$.
If $n$ is odd we can change type 0 to type II (and vice versa),
but this is possible only if $n$ is odd. Now we can state following rule:
{\it An orbifold string theory is type 0 or type II according to $\sum_i
k_i$ is even or odd. } For $C^r/\Z_n$, $r=1,2$, the result is consistent
with \cite{takayanagi} obtained from the partition function.

\subsection{ From partition function to GSO projection}

First consider $\C/\Z_n$ and $\C^2/\Z_n$.
By considering the low temperature limit of orbifold partition
functions\cite{lowe,takayanagi,dabholkar},
one can prove that the GSO projection is acting on chiral rings in the
following manner \cite{sin2}.
\begin{enumerate}
\item {\bf $\C/\Z_n$:} Let $q_l:=n\{k/n\}$ in the $c$ ring of
$n(k)$ theory, and $G_q=[lk/n]$. If $G_q$ is odd the
chiral GSO projection keeps $q$ in $c$ ring, project out $\bar q:=n-q$
in $a$ ring.
If $G_q$ is even, it keeps $\bar q$ in $a$ ring, project out $q$ in
$cc$ ring.

 \item {\bf $\C^2/\Z_n$:} Let $q_l:=(n\{k_1l/n\},n\{k_2l/n\})$ in the
$cc$ ring of
$n(k_1,k_2)$ theory, and $G_q=[lk_1/n]+[lk_2/n]$. If $G_q$ is odd the
chiral GSO projection keeps $q$ in $cc$ ring, project out $\bar q$ in
$ca$ ring.
If $G_q$ is even, it keeps $\bar q$ in $ca$ ring, project out $q$ in
$cc$ ring.
\end{enumerate}
For $\C^2/\Z_n$, this result is consistent with \cite{hkmm}.

{\bf Examples:}

\begin{enumerate}
 \item $n(1,1)$: $G=[j/n]+[j/n]=0$, hence all $cc$-ring and
$aa$-ring elements are projected out.
 All $ca$ and $ac$ ring elements survive under GSO.
 \item $n(1,-1)$: $G=[j/n]+[-j/n]=0+[-1+(n-j)/n]=-1$, hence all
$cc$-ring and $aa$-ring elements survive under GSO.
 All $ca$ and $ac$ ring elements are projected out.
 \item $n(1,n-1)$: $G=[j(n-1)/n]=[j-j/n]=j-1$. Hence, alternating.
 Surviving elements are $j=1$: (1,1); $j=2$: (2,n-2); $j=3$: (3,3);
etc.
 \item $n(1,1-n)$: $G=[j(1-n)/n]=[-j+j/n]=-j$: Alternating projection.
 Surviving elements are $j=1$: (1,1); $j=2$: (2,n-2); $j=3$: (3,3);
etc.
\end{enumerate}

From the examples above, it is quite obvious that the set of surviving
spectrum of $n(1,k)$ and that of $n(1,-k)$
are identical. The reason is because the $ca$ ring of $n(1,k)$ is the
same as the $cc$ ring of $n(1,-k)$ and this relation is
true even at the GSO projection. One can see this by simply calculating
the $G$ parity of $cc$ ring of each theory. For $n(1,k)$, $G=[jk/n]$ and
for $n(1,k)$, $G=[-jk/n]=-[jk/n]-1.$
They differ by one as desired. Therefore, two theories are isomorphic as
string theories.
On the other hand, $n(1,k)$ and $n(1,k-n)$ have the same spectrum before
GSO projection, but they are very different
after GSO projection.

Next we consider $\C^3/\Z_n$.
To derive the GSO rule, we need partition function and its limiting
behaviour.
Now let us calculate partition function $\C^3/\Z_n$.
Our calcualtion is based on \cite{tseytlin3}.
The relevant parts are summarized in the appendix A.
The partition function that we need is then
\be
Z=cV_3 R\int {d^2\tau \over \tau_2^2}\sum^\infty_{\omega,\omega'=-\infty}
e^{-{\pi \over {\alpha'\tau_2}}R^2(|\omega'-\tau\omega|^2)}
\times
{
{|\theta_1(\chi'|\tau)\theta_1(\chi_2'|\tau)\theta_1(\chi_3'|\tau)
\theta_1(\chi_4'|\tau) |^2 } \over
{| \theta_1(\chi_1|\tau)\theta_1(\chi_2|\tau)\theta_1(\chi_3|\tau)
\eta(\tau)^3|^2}
}.
\ee

Using Riemann's quartic identity listed in appendix C,
we can write GS partition function as RNS one.
Taking the orbifold limit($R\to 0$) as in \cite{takayanagi} $\C^3/\Z_n$
partition function becomes
\ba
&&Z(\tau)={1\over {4N}}\sum^{n-1}_{l,m=0}\left[
|\theta_3(\nu^1_{lm}|\tau)\theta_3(\nu^2_{lm}|\tau)
\theta_3(\nu^3_{lm}|\tau)\theta_3(0|\tau) \right. \no
&&\;\;\;\;\;\;\;\;\;\;\left.
-(-1)^{(k_1+k_2+k_3)\alpha}\theta_2(\nu^1_{lm}|\tau)
\theta_2(\nu^2_{lm}|\tau)\theta_2(\nu^3_{lm}|\tau)\theta_2(0|\tau)
\right. \no
&&\;\;\;\;\;\;\;\;\;\;\left. -(-1)^{(k_1+k_2+k_3)\beta}
\theta_4(\nu^1|\tau)
\theta_4(\nu^2_{lm}|\tau)\theta_4(\nu^3_{lm}|\tau)\theta_4(0|\tau)|^2
\right]
\no
&& \;\;\;\;\;\;\;\;\;\;\times
|\theta_1(\nu^1_{lm}|\tau)\theta_1(\nu^2_{lm}|\tau)
\theta_1(\nu^3_{lm}|\tau)\eta(\tau)^3|^{-2},
\ea
where
\be
\nu^i_{lm}={lk_i \over N}-{mk_i \over N}\tau.
\ee

By taking the low temperature($\tau_2 \to \infty \;or \;q\to 0$)
the partition function reduce the form
\be
Z \sim q^E,
\ee
where $E$ here represents minimal tachyon mass. $E$ depends on
whether $G= [{jk_1 \over n}]\pm[{jk_2 \over n}] \pm [{jk_3\over n}]$
is even or odd as well as on the ordering of $\{\mu_i\}$'s, the
fractional parts of $\mu_i:={jk_i\over n}$'s.
Here we wrote result  only and we gave more detail in appendix B.
The ordering of $\{\mu_i\}$'s gives 6 possibilities.
We first consider $\{\mu_1\} >\{\mu_2\}>\{\mu_3\}$ case.
For $G=even$, according to the range of
$\alpha:=\{\mu_1\}+\{\mu_2\}+\{\mu_3\}$ and
$\delta:=-\{\mu_1\}+\{\mu_2\}+\{\mu_3\}$, there are four possible cases:
\be
Z\sim \left\{%
\begin{array}{ll}
 q^0, & \hbox{$(0<\alpha<2, \delta>0)$;} \\
 q^{-\{\mu_1\} +\{\mu_2\} +\{\mu_3\}}, & \hbox{$(0<\alpha<2,
\delta<0)$;} \\
 q^{2-\{\mu_1\} -\{\mu_2\} -\{\mu_3\} }, & \hbox{$(2<\alpha<3,
\delta>0)$;} \\
 q^{2-2\{\mu_1\}}, & \hbox{$(2<\alpha<3, \delta<0)$.} \\
\end{array}%
\right.
\ee
The second and third cases give the spectrum of $acc$-ring, $aaa$-ring
respectively. In the first case, the spectrum is all marginal and this
is interesting, since we did not require any inequality like $\{\mu_1\} =
\{\mu_2\} +\{\mu_3\} $.
It means that marginal operator can be realized as a bulk in weight space
apart from the boundary of relevant and irrelevant regions.
However, since these states do not satisfy the charge-mass relation $h=q/2$, they are not  chiral states.  
The last case can not be realized since the $2<\alpha<3, \delta<0$  can
not be.
For $G$ odd, there are four cases according to the range of
$\alpha$ and $\beta:=\{ \mu_1 \}- \{ \mu_2 \} +\{ \mu_3 \}$.
\be
Z\sim
\left\{%
\begin{array}{ll}
 q^{\{\mu_1\} +\{\mu_2\} +\{\mu_3\} -1}, & \hbox{($0<\alpha\leq1$,
$0<\beta\leq 1$);} \\
 q^{2\{\mu_3\}} , & \hbox{$(0<\alpha\leq1, 1<\beta<2)$;} \\
 q^0, & \hbox{$(1<\alpha<3, 0<\beta\leq1)$;} \\
 q^{1-\{\mu_1\}-\{\mu_2\}+\{\mu_3\}}, & \hbox{$(1<\alpha<3,
1<\beta<2)$.} \\
\end{array}%
\right.
\ee
Here first and fourth cases give us the $ccc$ and $aac$ rings
respectively. The second case can not be realized and the third case
gives us a bulky range of marginal operators mentioned above.
In summary, for $\{\mu_1\} >\{\mu_2\}>\{\mu_3\}$ ordering,
we get $ccc,aac$ from  odd-$G$ and $acc,aaa$ rings from even $G$
and there are marginal regions.

Similarly, using the fact that the least $\mu_i$ give the $c$ in odd
cases and largest $\mu_i$ gives the $a$ in even cases, we can tabulate
the rings from each ordering.
\begin{table}
 \centering
\begin{tabular}{|c|c|c|}
 \hline
 ordering & $G$=even & $G$=odd \\
 \hline
 $\{\mu_1\}>\{\mu_2\}>\{\mu_3\}$ & $acc, aaa$ & $ccc,aac$ \\
 $\{\mu_1\}>\{\mu_3\}>\{\mu_2\}$ & $acc, aaa$ & $ccc,aca$ \\
 $\{\mu_2\}>\{\mu_3\}>\{\mu_1\}$ & $cac, aaa$ & $ccc,caa$ \\
 $\{\mu_2\}>\{\mu_1\}>\{\mu_3\}$ & $cac, aaa$ & $ccc,aac$ \\
 $\{\mu_3\}>\{\mu_2\}>\{\mu_1\}$ & $cca, aaa$ & $ccc,caa$ \\
 $\{\mu_3\}>\{\mu_1\}>\{\mu_2\}$ & $cca, aaa$ & $ccc,aca$ \\
 \hline
\end{tabular}
\caption{Chiral rings from various orderings}\label{ordering}
\end{table}
The GSO projection rule read off from the Table \ref{ordering} is as
follows:
\be
\left\{%
\begin{array}{ll}
 \;ccc,\;caa,\;aca,\;aac\; {\rm rings}: & \hbox{only
$G=\sum_i[lk_i/n]=odd$ cases survive.} \\
 \;aaa,\;acc,\;cac,\;cca\; {\rm rings}: & \hbox{only
$G=\sum_i[lk_i/n]=even$ cases survive.} \\
\end{array}%
\right.
\ee
 While it is largely by hand in other approaches, here the method is uniform and the result is simple.

\section{ Discussion and Conclusion}
In this paper we explicitly worked out the chiral rings and GSO
projection rules for non-compact orbifolds $\C^r/\Z_n$, $r=1,2,3$, which lead to the equivalence of NSR-GS formalisms.  We used  mode analysis to construct the chiral ring and
 derived and used partition functions of NSR formalism
 obtained from Green-Schwarz one transformed by the Riemann's theta identities.
The spectrum analysis from this partition function gives the surviving condition of individual energy values, which we identify as the rule of the GSO projection.
As a side remark, we found that there are unexpected rich spectrum of non-BPS marginal operators in $\C^3/\Z_n$ whose existence is not yet well understood from geometric point of view. We also mention that one of the main motive to discuss the GSO projection is to discuss the m-theorem \cite{sin3} in type II.  We wish to come back this issue in later publications.

\vskip.5in 
\noindent {\bf \large Acknowledgement} \\
This work is partially supported by the KOSEF 1999-2-112-003-5. 
We would like to thank A. Adams and T. Takayanagi for useful correspondences.

\appendix{Calculation of partition function }
Here we sketch the calculation of partition function of (NS,NS) Melvin with three magnetic parameters, $b_s(s=1,2,3)$. For more details for this section refer to \cite{tseytlin3}.
The three parameter solution satisfy the relation $\sum_s \pm b_s=0$, explicitly 
\be
b_1=b_2+b_3 \quad or\quad b_1=-(b_2+b_3)\quad or\quad b_1=b_2-b_3 \quad
or\quad b_1=-(b_2-b_3).
\ee
In this background the Lorentz group is broken to
\be
SO(8)\to SO(2)\times SO(2)\times SO(2)\times SO(2).
\ee
The three of $SO(2)$ represent rotations in the 1-2, 3-4, and 5-6
plane respectively. Fermion representation decompose as follows
\ba
&&\;\;\;\;\;S_R=\psi^{+++}_R \oplus\psi^{++-}_R\oplus \psi^{+-+}_R
\oplus \psi^{+--}_R\oplus \psi^{-++}_R \oplus \psi^{-+-}_R
\oplus \psi^{--+}_R \oplus \psi^{---}_R, \no
&&8_R\to (1_R,{1\over 2},{1\over 2},{1\over 2})\oplus
(1_R,{1\over 2},{1\over 2},-{1\over 2})\oplus
(1_R,{1\over 2},-{1\over 2},{1\over 2})\oplus
(1_R,{1\over 2},-{1\over 2},-{1\over 2})\oplus \no
&&\;\;\;\;\;(\bar{1}_R,-{1\over 2},{1\over 2},{1\over 2})\oplus
(\bar{1}_R,-{1\over 2},{1\over 2},-{1\over 2})\oplus
(\bar{1}_R,-{1\over 2},-{1\over 2},{1\over 2})\oplus
(\bar{1}_R,-{1\over 2},-{1\over 2},-{1\over 2}).
\ea
Bosonic Lagrangian has the form
\be
L_B=\sum_{s=1}^3 (\partial_+ +ib_s\partial_+ y)z_s
(\partial_- -ib_s \partial_- y)z^*_s,
\ee
where $z_s=x_s+ix_s$.
The fermionic Lagrangian is given by
\ba
L_F&=&i\bar{\psi}^{---}_R(\partial_+  +{i \over 2}(b_1+b_2+b_3)\partial_+
y)
\psi^{+++}_R + i \bar{\psi}^{-++}_R
(\partial_+ +{i \over 2}(b_1-b_2-b_3)\partial_+y )\psi^{+--}_R \no
&+& i\bar{\psi}^{--+}_R(\partial_+ +{i\over 2}(b_1-b_2+b_3)\partial_+ y)
\psi^{++-}_R + i \bar{\psi}^{+--}_R(\partial_+ + {i\over 2}(b_1+b_2-b_3)
\partial_+ y) \psi^{-++}_R  \no
&+&\;\; (R\to L).
\ea

The partition function becomes
\ba
&&Z(R,b_1,b_2,b_3)=cV_3 R\int {d^2\tau \over \tau_2^2}
\sum^\infty_{\omega,\omega'=-\infty}e^{-{\pi \over {\alpha'\tau_2}}
(R^2|\omega'-\tau\omega|^2)}Z_0(\tau,\bar{\tau};\chi_s,\bar{\chi}_s) \no
&&\times \left [
Y(\tau,\bar{\tau};{1\over 2}(\chi_1+\chi_2+\chi_3),{1\over
2}(\bar{\chi}_1
+\bar{\chi}_2+\bar{\chi}_3))Y(\tau,\bar{\tau};{1\over
2}(\chi_1-\chi_2-\chi_3),
{1\over 2}(\bar{\chi}_1-\bar{\chi}_2-\bar{\chi}_3)) \right. \no
&& \left. \times Y(\tau,\bar{\tau};
{1\over 2}(\chi_1-\chi_2+\chi_3),{1\over 2}(\bar{\chi}_1-
\bar{\chi}_2+\bar{\chi}_3))Y(\tau,\bar{\tau};{1\over 2}(
\chi_1+\chi_2-\chi_3),{1\over 2}(\bar{\chi}_1+\bar{\chi}_2-\bar{\chi}_3))
\right ] \no
&& \times \left [
Y(\tau,\bar{\tau};\chi_1,\bar{\chi}_1)Y(\tau,\bar{\tau};
\chi_2,\bar{\chi}_2)Y(\tau,\bar{\tau};\chi_3,\bar{\chi}_3) \right ]^{-1},
\ea
where $Y$ is given by
\be
Y(\tau,\bar{\tau};\chi,\bar{\chi})=e^{\pi(\chi-\bar{\chi})^2/2\tau_2}
{\theta_1(\chi|\tau) \over {\chi \theta'_1(0|\tau)}}
{\theta_1(\bar{\chi}|\tau)\over {\bar{\chi}\theta'_1(0|\bar{\tau})}}
\ee
We defined
\be
\chi_s=b_2 R (\omega' -\tau \omega),\quad
\bar{\chi}_s=b_s R (\omega'-\tau \omega),\quad
s=1,2,3.
\ee

Taking the limit $R\to 0$ and setting $b_sR={k_s\over N}$
lead us to orbifold partition function.

\appendix{Low temperature limit of the partition function}
Using the identity
\ba
&&\theta_1( {\nu^1_{lm}\pm\nu^2_{lm}\pm \nu^3_{lm} \over 2}|\tau) \no
&&=i \sum(-1)^n q^{{1\over 2}(n-{1\over 2}-{1\over 2}
(\mu_1\pm \mu_2\pm\mu_3))^2 } q^{-{1\over 8}(\mu_1\pm\mu_2\pm\mu_3)^2}
e^{i\pi (n-{1\over 2}){l\over n}(k_1 \pm k_2\pm k_3)},
\ea
the relevent part of partition function of type II can be written as
$Z\sim q^{N-D}$
\be
N=\left\{%
\begin{array}{ll}
 \(\half-\frac{\alpha}{2}\)^2, & \hbox{if $0<\alpha<2$} \\
 \({3\over2}-\frac{\alpha}{2}\)^2, & \hbox{if $2<\alpha<3$} 
\end{array}%
\right\} + \(\half-\frac{\beta}{2}\)^2 +\(\half-\frac{\gamma}{2}\)^2
+\left\{%
\begin{array}{ll}
\(\half-\frac{\delta}{2}\)^2, & \hbox{if $0<\delta$} \\
 \({\half}-\frac{\delta}{2}\)^2, & \hbox{if $0>\delta$} 
\end{array}%
\right\}
\ee
for even $G$,
\be
N=\left\{%
\begin{array}{ll}
 \(\frac{\alpha}{2}\)^2, & \hbox{if $0<\alpha<1$} \\
 \({1}-\frac{\alpha}{2}\)^2, & \hbox{if $1<\alpha<3$} 
\end{array}%
\right\}  + \(\frac{\beta}{2}\)^2 +\(\frac{\delta}{2}\)^2
+\left\{%
\begin{array}{ll}
 \(\frac{\gamma}{2}\)^2, & \hbox{if $0<\gamma<1$} \\
 \(1-\frac{\delta}{2}\)^2, & \hbox{if $0<\gamma<2$} 
\end{array}%
\right\}
\ee
for odd $G$, and
\be
D=\sum_i\{\mu_i\}^2-\sum_i\{\mu_i\}+1,
\ee
where
$\alpha:=\{\mu_1\}+\{\mu_2\}+\{\mu_3\}$,
$\beta:=\{\mu_1\}-\{\mu_2\}+\{\mu_3\}$,
$\gamma:=\{\mu_1\}+\{\mu_2\}-\{\mu_3\}$ and
$\delta:=-\{\mu_1\}+\{\mu_2\}+\{\mu_3\}$.
Considering all these cases one can obtain the 4 cases of $G=even$ four
cases of $G=odd$ for the ordering $\mu_1>\mu_2>\mu_3$. For other
orderings calculations are similar.

\appendix{Theta function  identities }

The partition function in the Green-Schwarz(GS) formalism
can be easily transformed to Ramond-Neveu-Schwarz(RNS) one
by using Riemann theta identity
\be
2\theta_1^4(\nu|\tau)=\theta_3^4(\nu|\tau)-\theta_2^4(\nu|\tau)
-\theta_4^4(\nu|\tau)+\theta_1^4(\nu|\tau).
\ee
More generally,  we have
\be
2\prod_{a=1}^4\theta_1(\nu_a'|\tau)
=\prod_{a=1}^4\theta_3(\nu_a|\tau)-\prod_{a=1}^4\theta_2(\nu_a|\tau)
-\prod_{a=1}^4\theta_4(\nu_a|\tau)+\prod_{a=1}^4\theta_1(\nu_a|\tau),
\ee
where
\ba
&&2\nu_1' =\nu_1+\nu_2+\nu_3+\nu_4,\;\;\;2\nu_2' =\nu_1+\nu_2
-\nu_3-\nu_4, \no
&&2\nu_3'=\nu_1-\nu_2+\nu_3-\nu_4,\;\;\;2\nu_4'=\nu_1-\nu_2-\nu_3+\nu_4.
\label{nu}
\ea
We need the definition and a property of $\theta_1$:
\be
\theta_1(\nu|\tau)=2q^{{1\over 8}}\sin(\pi \nu)\prod^\infty_{n=1}
(1-q^n)(1-e^{2\pi i \nu}q^n)(1-e^{-2\pi i \nu}q^n)
=i\sum^\infty_{n=-\infty}(-1)^nq^{{1\over 2}(n-{1\over 2})^2}
z^{n-{1\over 2}}.\ee

\appendix{Chiral Rings and Enhanced (2,2) SUSY }

Here, we will show that for $\C^2/\Z_n$, any worldsheet
fermion generated tachyon can be constructed as a BPS state, i.e., a
member of a chiral ring.
Essential ingredient is the existence of the 4 copies of (2,2) worldsheet
SUSY for this special theory.

Characterizing a state as a chiral or anti-chiral state gives an
extremely powerful
result since we can utilize the (2,2) worldsheet supersymmetry.
If all the tachyon spectrum are chiral or anti-chiral, the analysis of
the tachyon condensation could be much easier.
However, in reality it is not the case.
For example, when $a_2<\half<a_1$, $\psi^*_1|0> \sim
u_1^{na_1}u_2^{na_2}$ and $\psi_2|0>\sim {\bar u}_1^{n(1-a_1)}
{\bar u}_2^{n(1-a_1)}$ are chiral and anti-chiral state
respectively, while $|0>$ and $\psi^*_1\psi_2|0>$ are neither of
them.\footnote{One may argue that we have not considered the left-right
combination and
it might be such that
left-right combination non BPS tachyon might be projected out. However,
examining the low temperature behavior of the partition
function\cite{sin2}, we can easily see that the string theory does
contain a tachyon with
$\frac{1}{4}\alpha'M^2=-\half|a_1-a_2|$ as well as
$\frac{1}{4}\alpha'M^2=-\half|a_1+a_2-1|$.
In fact, since we are looking for lowest tachyonic spectrum which comes
from (NS,NS)
sector the level matching condition requires that
$\Delta_L=\Delta_R$ and we do not get $-|a_1-a_2|$ from the
(chiral,chiral) or (anti-chiral,anti-chiral) states.
That is, those spectrum with mass of the form
$\frac{1}{4}\alpha'M^2=-\half|a_1-a_2|$
is in fact not a SUSY state according to our definition of (2,2) SUSY.
For the level matching between left NS and right R
sectors, we need to consider the modular invariance that leads to
GSO projection $n(E_{NS}-E_{R})=0$ mod 1 \cite{lowe}.
Even in the case we combine left chiral and right anti-chiral, we do not
get
the spectrum of type $\frac{1}{4}\alpha'M^2=-\half|a_1-a_2|$.
}
This issue is particularly relevant in case
the lowest mass in the given twisted sector is neither chiral nor
anti-chiral.
\footnote{One example is the 10(1,3) theory.}

However, we will see that one can improve the situation by recognizing
that there
are enhanced SUSY in orbifold thoeries.
We will show that all twisted sector tachyons generated by world sheet
fermions
can be considered as chiral states by redefining the generator of the
supersymmetry algebra.
For this, let's define
$L^{(i)},G^{(i)\pm},J^{(i)}$ as the generators of ${\cal N}=2$
superconformal algebra
in $i$-th complex plane. Usually we define $L=L^{(1)}+L^{(2)}$,
$J=J^{(1)}+J^{(2)}$ and $G^{+}=G^{(1)+}+G^{(2)+}$ and the last was used
above to identify the chiralities. However, it is a simple matter
to check that we can also define the ${\cal N}=2$ superconformal algebra
by defining $G^{+}=G^{(1)+}+G^{(2)-}$ with corresponding change in
$J=J^{(1)}-J^{(2)}$ but the same $L=L^{(1)}+L^{(2)}$. We call this
$(+-)$ choice of $G^+$ as $G^+_{ca}$, while we call the previous (++)
choice as $G^+_{cc}$. The fact that we need to change the sign of
$J^{(2)}$
means that we need to count the $U(1)$ charge of $u_2, {\bar u_2}$ as
$-1, +1$ respectively
while $u_1, {\bar u_1}$ as $1, -1$ as before.
The choice of $G^+$ corresponds to the target space complex structure.
This phenomena is due to the special geometry
of target space in which each complex plane have independent complex
structure so that
to define a complex structure of the whole target space, we need to
specify one
in each complex plane.

Since $J\sim \psi^*\psi$ and $G^+ \sim \psi^*\p X$ and $G^- \sim
\psi\p X^*$, the above change of generator construction
corresponds to the change in the complex structure in the target
space, i.e, interchanging starred fields and un-starred fields with
the notion of positivity of charge also changed: $\psi^*$ has $-1$
charge and $\psi$ has $+1$ charge, which is opposite to the
previous case.

Since the chirality is defined by this new choice of $G^+$, we
now have different classification of tachyon states: for
example, in $a_2<\half<a_1<1$ case, $\psi^*_1\psi_2|0> \sim
u_1^{na_1}{\bar u}_2^{n(1-a_2)}$ and $|0> \sim {\bar
u}_1^{n(1-a_1)}u_2^{na_2}$ are chiral and anti-chiral state
respectively. Notice that they were neither $chiral$ nor
$anti-chiral$ under $G^+_{cc}$. On the other hand,
$\psi^*_1|0> \sim u_1^{na_1}u_2^{na_2}$ and $\psi_2|0>\sim
{\bar u}_1^{n(1-a_1)}{\bar u}_2^{n(1-a_1)}$ are neither chiral
nor anti-chiral in the new definition of $G^+$.
Similarly, we can classify other parameter zones. The result can be
conceptually summarized as follows: for $G_{cc}$, $G_{ca}$, $G_{ac}$,
$G_{aa}$
the monomial basis of local chiral ring is
generated by $u_1^{k_1}u_2^{k_2}$, $u_1^{k_1}{\bar u}_2^{n-k_2}$,
${\bar u}_1^{n-k_1}u_2^{k_2}$ and ${\bar u}_1^{n-k_1}{\bar
u}_2^{n-k_2}$ respectively, while the anti-chiral ring is
generated by ${\bar u}_1^{n-k_1}{\bar u}_2^{n-k_2}$, ${\bar
u}_1^{n-k_1}u_2^{k_2}$, $u_1^{k_1}{\bar u}_2^{n-k_2}$,
$u_1^{k_1}u_2^{k_2}$ respectively.
Notice that anti-chiral ring of $G_{cc}$ is chiral ring of $G_{aa}$,
while
anti-chiral ring of $G_{ca}$ is chiral ring of $G_{ac}$.
Therefore we may consider only chiral ring of each complex structure.
We call the chiral ring of $G_{cc}$ complex structure as ${cc}$-ring.
We define $ca$-ring, $ac$-ring and $aa$-ring similarly.

\begin{figure}[htbp2]
\epsfysize=5cm
\centerline{\epsfbox{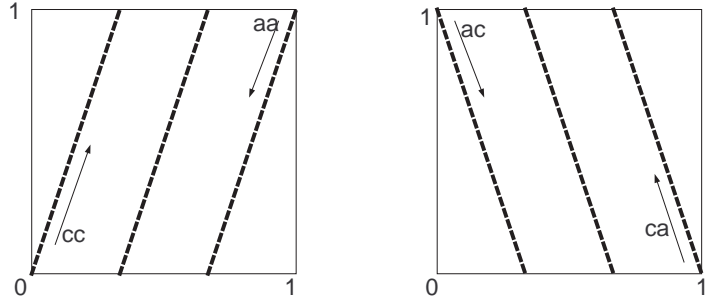}}
\caption{ \scriptsize Weight points of $cc,aa$ and $ac,ca$ rings in
weight space.
x- and y-axis represent
$2\Delta_1(j)$ and $2\Delta_2(j)$. Arrows represent the
direction and starting point of corresponding ring as $j$ increases
from 1 to n-1.
Plot is drawn for $k_1=1,\; k_2=3$.
}
 \label{ccNac}
\end{figure}

It is convenient to consider the weight of a state as sum of contribution
from each complex plane.
For example, the weight of $u_1^{na_1}u_2^{na_2}$ can be considered as
sum of $a_1$ from $u_1$
and $a_2$ from $u_2$. $(a_1,a_2)$ form a point in the weight space.
As we vary $j$ in $a_i=\{jk_i/n\}$,
the trajectory of the point in weight space will give us a parametric
plot in the plane.
In the figure \ref{ccNac}, we draw for weight points of $cc$ and $aa$
rings in the first figure
and those of $ca$ and $ac$ rings in the second figure of Fig.
\ref{ccNac}.
In order to compare these spectrum with $a_1$ and/or $a_2$ negative
cases,
we work out the weight of the states in Table \ref{cplx}.
\begin{table}
 \centering
\begin{tabular}{|c||c|c||c|c|} \hline
 $ G$ & $c$-ring & $2\Delta$ & $a$-ring & $2\Delta$
 \\ \hline \hline
$ G_{cc}$ & $u_1^{na_1}u_2^{na_2}$ & $a_1+a_2$ & ${\bar
u_1}^{n(1-a_1)}{\bar u_2}^{n(1-a_2)}$ & $2-a_1-a_2$
 \\ \hline
$ G_{ca}$ & $ u_1^{na_1}{\bar u}_2^{n(1-a_2)}$ & $ a_1-a_2+1$ & $ {\bar
u_1}^{n(1-a_1)}{ u_2}^{na_2}$ & $ 1-a_1+a_2$
 \\ \hline
$ G_{ac}$ & $ {\bar u}_1^{n(1-a_1)}u_2^{na_2}$ & $ 1-a_1+a_2$ & ${
u_1}^{na_1}{\bar u_2}^{n(1-a_2)}$ & $ a_1-a_2+1$
 \\ \hline
$G_{aa}$ & $ {\bar u}_1^{n(1-a_1)}{\bar u}_2^{n(1-a_2)}$ & $ 2-a_1-a_2$
& ${u_1}^{na_1}{u_2}^{na_2}$ & $ a_1+a_2$
 \\ \hline
\end{tabular}
 \caption{Monomial basis of chiral and anti-chiral rings and their
weights
 for various choices of target space complex structures. }\label{cplx}
\end{table}
By comparing the two Table \ref{sign} and Table \ref{cplx}, it is clear
that the spectrum of
$ca$-ring of $n(k_1,k_2)$ theory is equal to the $cc$-chiral ring of
$n(k_1,-k_2)$ theory.
So the change in complex structure $u_i\to {\bar u_i}$ is equivalent to
the change in generator
$k_i\to -k_i$ keeping the complex structure fixed. For string
theory, we have to consider all four different complex structures.
That is, we may consider 4 sets of spectra generated by
$(k_1,k_2)$, $(-k_1,k_2)$, $(k_1,-k_2)$ and $(-k_1,-k_2)$ all
together.

Summarizing, we have shown that any of the lowest
tachyon spectrum generated by the worldsheet fermions, can be considered
as chiral state
by choosing a worldsheet SUSY generator
appropriately; any of them  belongs to one of 4 classes: $cc$-, $ca$-,
$ac$-, $aa$- ring depending on the choice of
complex structure of $\C^2$.
This is explicit in the Table \ref{allring}.
\begin{table}
 \centering
\begin{tabular}{|c|c|c|c|c|}\hline
($a_1-\half,a_2-\half$) & $cc$ & $ca$ & $ac$& $aa$ \\ \hline
 $--$ & $|0>$ & $\psi_2|0>$ & $\psi_1|0>$& $\psi_1\psi_2|0>$ \\ \hline
 $-+$ & $\psi_2^*|0>$ & $|0>$ & $\psi_1\psi_2^*|0>$ & $\psi_1|0>$ \\
\hline
 $+-$ & $\psi_1^*|0>$ & $\psi_1^*\psi_2|0>$ & $|0>$& $\psi_2|0>$ \\
\hline
 $++$ & $\psi_1^*\psi_2^*|0>$ & $\psi_1^*|0>$ & $\psi_2^*|0>$& $|0>$
\\ \hline
 \end{tabular} \\
\caption{ For a given twisted sector, any tachyon generated by
worldsheet fermions is
an element of one of the 4 possible chiral rings.}\label{allring}
\end{table}
We emphasize that these chiral rings do not co-exist at the same time.
For example, when $cc$-ring is active ( chosen),
then $aa$-ring exists as its anti-chiral ring and other two are not
chiral or anti-chiral ring. But for our purpose, for any tachyon state,
there is a choice of
complex structure in which the given state is a chiral primary.
For example, if a tachyon in $ca$ ring is condensed, the spectrum of
entire $ca$-ring is 
well controlled by the worldsheet supersymmetries generated by  
$G_1^{+},G_2^-$.
As a consequence, we will be able to calculate the fate of those
controlled spectrum.
This is powerful since if we know that initial and final thoeries are
described by an orbifold theories
\cite{aps,vafa,hkmm}, knowing those of a few spectrum completely fixes
entire tower
of the string spectrum in the final theory.
The same phenomena arise for all $\C^r/\Z_n$. Any worldsheet fermion
generated
tachyon state is a chiral primary by properly choosing the target space
complex structure among
$2^r$ possibilities defined by the $\sum_{i=1}^r G_i^\pm$.
There are $2^r$ $(2,2)$ world sheet supersymmetries instead of one.
\footnote{The notion of enhanced symmetry already appeared in literature
implicitly.
For example in \cite{hkmm,kdecay}, the notion of $cc,ca$ ring is
discussed and
the chiral ring elements were described in terms of bosonization. }
In fact this happens for any tensor product of ${\cal N}=2$ SCFT's.

We end this subsection with a few comments.
\begin{itemize}
\item
The weight space is a lattice in torus of size $n\times n$.
The identification of weights by modulo $n$ corresponds to shifting
string modes.
However, periodicity of the generator $(k_1,k_2)$ is $2n$ and
$(k_1,k_2)$ and $(k_1,k_2+n)$ do not generate the same theory in
general.
We choose the standard range of $k_i$ between $-n+1$ to $n-1$.
This is because the GSO projection depends not only on the R-charge
vector $(\{jk_1/n\},\{jk_2/n\})$ but also on the $G$-parity number
$G=[jk_1/n]+[jk_2/n]$.
We will come back to this when we discuss the GSO projection.

\item
When $n$ and $k_i$ are not relatively prime,
we have a chiral primary whose R-charge vector is $(p/n,0)$.
We call this as the reducible case and eliminate from our interests.
This is a spectrum that is not completely localized at the tip of the
orbifold.
Sometimes, even in the case we started from non-reducible theory, a
tachyon condensation leads us to
the reducible case.
\end{itemize}


\end{document}